\documentclass[prl,twocolumn]{revtex4}
\usepackage{amsmath}
\usepackage{dcolumn}
\usepackage{bm}
\usepackage{epsfig}
\usepackage{epstopdf}
\usepackage{graphicx}
\usepackage{amsfonts}
\usepackage{amssymb}
\usepackage{mathrsfs}
\usepackage{color}
\usepackage{multirow}
\usepackage{mathbbol}
\usepackage{bbold}
\usepackage{hyperref}
\usepackage[usenames,dvipsnames]{xcolor}
\usepackage{float}
\usepackage{bbm}
\usepackage{pifont}
\emergencystretch=\maxdimen
\hypersetup{
	colorlinks = true,
	linkcolor = blue,
	anchorcolor = blue,
	citecolor = blue,
	filecolor = blue,
	urlcolor = blue
}
\renewcommand{\footnoterule}{  \hrule width \textwidth height 1pt
	\kern 2pt}
	
\begin{document}

\title{Single-Atom Verification of the Optimal Trade-Off between Speed and Cost in Shortcuts to Adiabaticity}
\author{J.-W. Zhang$^{1}$} \thanks{Co-first authors with equal contribution}
\author{J.-T. Bu$^{2,3}$}\thanks{Co-first authors with equal contribution}
\author{J. C. Li$^{4,1}$}\thanks{Co-first authors with equal contribution}
\author{Weiquan Meng$^{5}$}\thanks{Co-first authors with equal contribution}
\author{W.-Q. Ding$^{2,3}$}
\author{B. Wang$^{2,3}$}
\author{W.-F. Yuan$^{2,3}$}
\author{H.-J. Du$^{2,3}$}
\author{G.-Y. Ding$^{2,3}$}
\author{W.-J. Chen$^{2,3}$}
\author{L. Chen$^{2,1}$}
\author{F. Zhou$^{2,1}$}
\email{zhoufei@wipm.ac.cn}
\author{Zhenyu Xu$^{5}$}
\email{zhenyuxu@suda.edu.cn}
\author{M. Feng$^{1,2,6}$}
\email{mangfeng@wipm.ac.cn}
\affiliation{$^{1}$Research Center for Quantum Precision Measurement, Guangzhou Institute of Industry Technology, Guangzhou, 511458, China \\
	$^{2}$State Key Laboratory of Magnetic Resonance and Atomic and Molecular Physics,
	Wuhan Institute of Physics and Mathematics, Innovation Academy of Precision Measurement Science and Technology, Chinese Academy of Sciences, Wuhan, 430071, China\\
	$^{3}$School of Physics, University of the Chinese Academy of Sciences, Beijing 100049, China \\
    $^{4}$Guangzhou Institute of Industrial Intelligence, Guangzhou, 511458, China \\
    $^{5}$School of Physical Science and Technology, Soochow University, Suzhou 215006, China \\
	$^{6}$Department of Physics, Zhejiang Normal University, Jinhua 321004, China }

\begin{abstract}
The approach of shortcuts to adiabaticity enables the effective execution of adiabatic dynamics in quantum information processing with enhanced speed. Owing to the inherent trade-off between dynamical speed and the cost associated with the transitionless driving field, executing arbitrarily fast operations becomes impractical. To understand the accurate interplay between speed and energetic cost in this process, we propose theoretically and verify experimentally a new trade-off, which is characterized by a tightly optimized bound within $s$-parameterized phase spaces. Our experiment is carried out in a single ultracold $^{40}$Ca$^{+}$ ion trapped in a harmonic potential. By exactly operating the quantum states of the ion, we execute the Landau-Zener model as an example, where the quantum speed limit as well as the cost are governed by the spectral gap. We witness that our proposed trade-off is indeed tight in scenarios involving both initially eigenstates and initially thermal equilibrium states. Our work helps understanding the fundamental constraints in shortcuts to adiabaticity and illuminates the potential of under-utilized phase spaces that have been traditionally overlooked.
\end{abstract}
\maketitle

Adiabatic dynamics requires an exceptionally slow evolution of quantum states, posing challenges for practical applications of quantum technology, which are strongly restricted by decoherence time. Fortunately, shortcuts to adiabaticity offers a pathway to achieving identical outcomes through expedited processes \cite{RMP2019}. Transitionless quantum driving exemplifies such techniques \cite{JPCA2003,JPCB2005,JPA2009}, whose core principle is to involve an ancillary counterdiabatic field $H_1 (t)$ within $t \in (0,\tau)$. Under the combined Hamiltonian $H(t)=H_0 (t)+H_1 (t)$, the fast quantum dynamics aligns with the adiabatic process governed solely by $H_0 (t)$. This quantum technology has been extensively employed to enhance the performance of elementary operations in quantum information science \cite{PRLXA,NPMM,Alonso2012,PRLA,PRLJJ,Paternostro2015,NCSD,NCYZ,Takahashi2017,NJPJA,PRLJA,PRLHY,PRLYW}, and applied to other research fields, such as quantum thermodynamics~\cite{PRLSS,PRLKJ,Muga2017,NJPZT,Paternostro2020,Alipour2021,PRLGJ,PRLAE}, quantum metrology~\cite{ECG} and even the population engineering of organisms in medical biology~\cite{NPSE}.

One natural question: what is the maximum speed achievable by such transitionless driving? Given the inherent trade-offs, a quantum speed limit (QSL) necessarily constrains the rapidity of this transitionless driving.
A seminal relationship between the cost rate and evolution speed has been established \cite{PRLSS}, indicating that instantaneous manipulation remains elusive owing to the requirement of an infinite cost rate.
However, establishing a direct relationship between the dynamical speed and the driving cost often encounters obstacles, leading to the prevalent use of the upper bound of dynamical speed, namely the QSL, to associate with the cost rate. Thus, the tightness of this bound and its ability to accurately reflect the true dynamical process emerge as two critical factors in assessing the precision of the trade-off between speed and cost.

Experimentally exploring such relationship is imperative and typical advancements of QSL experiments so far include detection of speedup in optical cavity QED systems \cite{Deffner2015PRL}, observation crossover between Mandelstam-Tamm and Margolus-Levitin bounds using fast matter wave interferometry \cite{Ness2021SA}, navigation of quantum brachistochrones between distant states of an atom across optical lattices \cite{Alberti2021PRX}, and execution of logical gates at the QSL in superconducting transmons \cite{Gong2023PRR}.

In this Letter, we propose theoretically and also verify experimentally a new trade-off between speed and cost, based on an optimized QSL bound using the $s$-parametrized phase space technique \cite{PRAWZ}. This approach yields a tight bound compared to those derived in Wigner phase space or Hilbert space. We demonstrate that this new trade-off surpasses the one outlined in \cite{PRLSS} that actually lacks tightness \cite{SM}. We also justify our findings experimentally in an ion trap platform using the well-known Landau-Zener (LZ) model, which is closely related to the Ising model \cite{PRLJ,PRLAM,PRBJM} and thus useful for quantum annealing \cite{JPCSA}. In this context, our present study substantially enhances the understanding of fundamental constraints governing the speed of quantum operations.
Notably, the tight QSL bound we investigate here outperforms the traditionally employed Wigner, Glauber-Sudarshan, and Husimi phase spaces. This finding strongly supports the notion that rarely used phase spaces may offer superior performance in certain quantum tasks.


To clarify this new trade-off, we first elucidate briefly the theory of the cost rate, quantum dynamical speed and limit. Consider a time-dependent Hamiltonian $H_0(t)$, characterized by its instantaneous eigenvalues $\{\epsilon_n(t)\}$ and eigenstates $\{\left| n_t \right\rangle\}$. In ideal situation, the adiabatic dynamics takes an infinitely long time and would not induce transitions between the eigenstates. In contrast, for a swift evolution in transitionless quantum driving without eigenstate transitions, it is necessary to apply an ancillary counterdiabatic field  with the form of $H_1(t)=i \hbar\left[\partial_t\left(\left|n_t\right\rangle\left\langle n_t\right|\right),\left|n_t\right\rangle\left\langle n_t\right|\right]$ \cite{RMP2019}.

A suite of cost rate functions for transitionless driving has been presented~\cite{PRAYS}. Notably, the most elementary among them, when disregarding the setup constant, takes the following subsequent form~\cite{PRLSS} as
\begin{equation}
\partial_tC =  ||H_{1}\left(t\right) ||=\sqrt{2}\hbar\sqrt{\left\langle\partial_t n_t \mid \partial_t n_t\right\rangle}, \label{Eq2a}	
\end{equation}
where $||X||=\sqrt{\operatorname{tr}\left(X^{\dagger} X\right)}$ denotes the Frobenius norm of the operator $X$ \cite{BookMA}.

In the study of dynamical speed, it is commonly characterized by the time derivative of a metric that denotes the overlap or disparity between an initial and its subsequent evolved state \cite{Note}. In this Letter, we utilize the relative purity \cite{QICAM}, a metric known for its computational efficiency, denoted by $P_t \left(\rho_0, \rho_t\right)=\operatorname{tr}(\rho_0 \rho_t)/\operatorname{tr}\left(\rho_0^2\right)$ \cite{PRLAI} or simply $ P_t \left(\rho_0, \rho_t\right)=\operatorname{tr}(\rho_0 \rho_t)$ \cite{Lidar98,PRLBA}. Throughout this work, we will adopt the latter to assess the QSL. 
However, the upper bound of evolution speed determined via such relative purity is not tight \cite{SM}. Fortunately, this limitation can be effectively circumvented through the application of the recently developed $s$-parameterized phase space technique in the context of QSL \cite{PRAWZ}.

The relative purity in $s$-parameterized phase space is expressed as $P_t\left(\rho_0, \rho_t\right)=\int d \mu(\eta) F_{\rho_0}^{-s}(\eta) F_{\rho_t}^s(\eta)$ \cite{PRAWZ}, where $F_A^s(\eta)$ is the phase space function of operator $A$, $\eta$ is a point in a phase space that determines the state $|\eta\rangle(\eta \rightarrow|\eta\rangle)$ in Hilbert space, and the index $s$ labels a family of Cahill-Glauber phase spaces \cite{Glauber1969a,Glauber1969b}. To acquire an optimal bound, the changing rate of the relative purity, namely the dynamical speed, is bounded by
\begin{equation}
	\left|\dot{P}_t\left(\rho_0, \rho_t\right)\right| \leq \sup_{s\in {S}}V_{\mathrm{QSL}}^s(t),
	\label{Eq3}	
\end{equation}
where $V_{\mathrm{QSL}}^s(t)=[\int d \mu(\eta) F_{\rho_0}^{-s}(\eta)^2 \int d \mu(\eta') |\dot F_{\rho_t}^{s}(\eta')|^2]^{1/2}$ \cite{PRAWZ} represents the QSL in Cahill-Glauber $s$-parametrized phase space, and $\mathrm{sup}$ signifies the least upper bound, related to the phase space parameter $s$ within a set $S$. Notably, $s=0, 1$, and $-1$ correspond to the well-known Wigner, Glauber-Sudarshan, and Husimi phase spaces, respectively \cite{Runeson2020,Rundle2021Review}.

Without loss of generality, let us consider $N$-level quantum systems. The least upper bound, i.e., Eq. (\ref{Eq3}), is achieved when $s\rightarrow -\infty$ and the tightness of such bound is ensured by the condition of
\begin{equation}
    b_{\mu}(0) \propto h_{\nu}(t) b_{\lambda}(t)f_{\nu \lambda \mu},\label{Eq25}
\end{equation}
where $b_\mu(t)=2 \operatorname{tr}\left(\rho_t T_\mu\right)$, $h_\mu(t)=2 \operatorname{tr}\left(H(t) T_\mu\right)$, $f_{\nu \lambda \mu}$ are totally antisymmetric regarding the interchange of any pair of its indices, and $T_\mu$ are the generators of SU($N$) Lie algebra. The summation over repeated Greek indices is implicit.

\textit{Proof.} To prove the condition stated in Eq. (\ref{Eq25}) for ensuring the tightness of the QSL upper bound in Eq. (\ref{Eq3}), it is essential to demonstrate that $F_{\rho_0}^{-s}(\eta)$ and $\dot{F}_{\rho_t}^s(\eta)$ are linearly dependent. For $N$-level quantum systems, we have $F_{\rho_0}^{-s}(\eta)=1/N+2 b_{\mu}(0) r_{-s} R_{\mu}$ and $\dot F_{\rho_t}^s(\eta)=(2/\hbar)r_s b_{\mu}(t)h_{\nu}(t)R_{\lambda}f_{\nu \mu \lambda}$ \cite{SM}, where $r_s=\sqrt{(N+1)^{1+s}} /2$ ($s\in\mathbbm{R}$), and $R_\mu=\left\langle\eta\left|T_\mu\right| \eta\right\rangle$. As $s$ approaches $-\infty$, the value of $r_{-s}$ tends towards infinity. Consequently, the first term, $1/N$, in $F_{\rho_0}^{-s}(\eta)$ becomes negligible and can be omitted. Thus, we can solve the set of simultaneous equations, i.e., Eq. (\ref{Eq25}) to ensure linear relevance between $F_{\rho_0}^{-s}(\eta)$ and $\dot F_{\rho_t}^s(\eta)$, and thereby guaranteeing the tightness of the upper bound in Eq. (\ref{Eq3}). More details can be found in \cite{SM}. \hfill $\blacksquare$

In this Letter, we focus on two-level (qubit) systems for experimental verification. $V_{\mathrm{QSL}}^s(t)$ in Eq. (\ref{Eq3}) is simplified as \cite{PRAWZ}
\begin{equation}
    V_{\mathrm{QSL}}^s(t)=\frac{1}{2 \hbar} \sqrt{3^s+|\vec{b}_0|^2}|\vec{h}_t \times \vec{b}_t|,\label{Eq4}
\end{equation}
where $\vec{b}_t=\operatorname{tr}(\rho_t \vec{\sigma})$ and $\vec{h}_t=\operatorname{tr}(H(t) \vec{\sigma})$. When the initial state is pure, we have $\Delta E=\frac{1}{2} |\vec{h}_t \times \vec{b}_t|$ and $\left|\vec{b}_0\right|^2=1$. Thus, the $s$-parametrized QSL bound of the qubit system in Eq. (\ref{Eq4}) is rewritten as $V^s_{\mathrm{QSL}}(t)=\sqrt{1+3^s} \Delta E/\hbar$, which is a Mandelstam-Tamm type bound \cite{MTbound}. In certain specialized scenarios, such as a qubit system initially prepared in one of the eigenstates of $H_0(0)$, we may discern analytically the relationship between the speed limit and the cost rate. After some algebra, Eq. (\ref{Eq4}) becomes $V_{\mathrm{QSL}}^s(t)=\sqrt{3^s+1}\sqrt{\left\langle\partial_t n_t \mid \partial_t n_t\right\rangle}$,
which leads to a new concise trade-off between the QSL and the cost rate
\begin{equation}
	\frac{V_{\mathrm{QSL}}^s(t)}{\partial_tC}=\frac{\sqrt{3^s+1}}{\sqrt{2}\hbar},
	\label{Eq7}
\end{equation}
governed by the $s$ parameters in phase spaces. Clearly, the optimal trade-off between the true dynamical speed and cost rate is upper bounded by $\left|\dot{P}_t\right| / \partial_t C \leq V_{\mathrm{QSL}}^{-\infty}(t)/\partial_t C=1/\sqrt{2}\hbar$ \cite{explain}.

\begin{figure}[t]
	\centering
	\includegraphics[width=8.6cm]{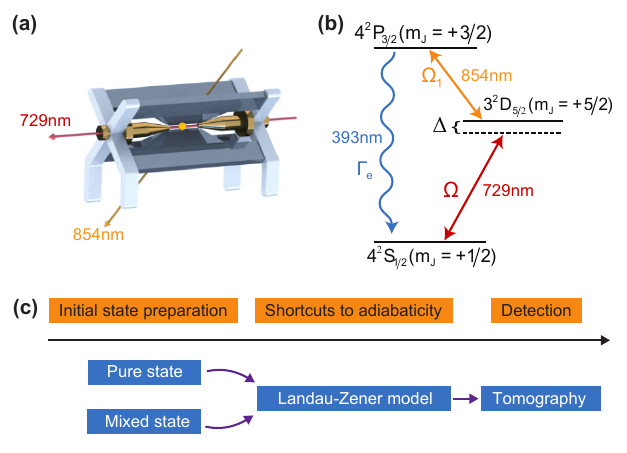}
	\caption{(a) The ion trap system, where the 729-nm laser beam is in parallel with the axial direction, and the 854-nm laser irradiates with an angle $45^\circ$ with respect to the axial direction. (b) Level scheme of the $^{40}$Ca$^+$ ion, where the double-sided arrows and the wavy arrow represent the laser irradiation and dissipation, respectively. (c) The experimental steps starting from the initialization to the final detection. }
	\label{Fig1}
\end{figure}

\begin{figure*}[t]
	\centering
	\includegraphics[width=18cm]{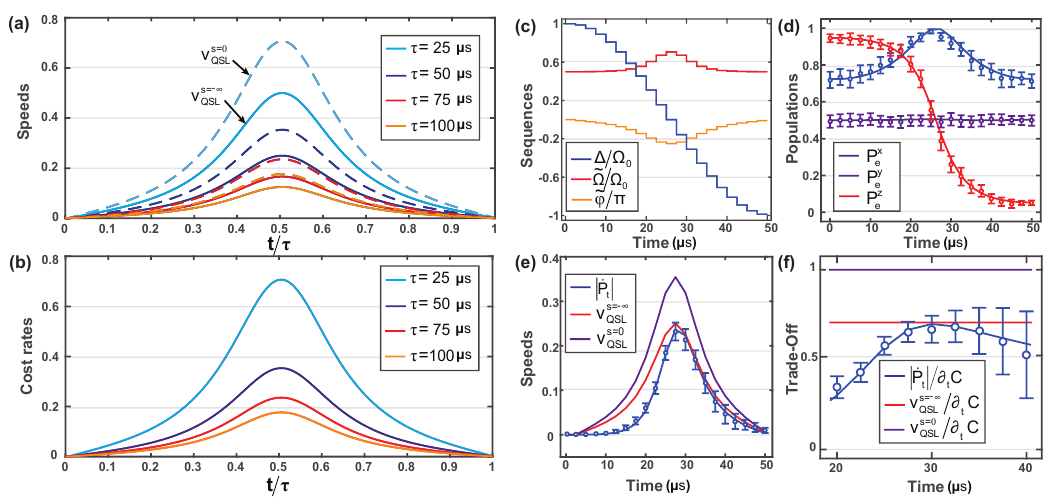}
	\caption{Trade-off between speed and cost in shortcuts to adiabaticity for an initially prepared eigenstate. Panels (a, b) display theoretical analyses over varying time periods from $\tau=25$ $\mu$s to 100 $\mu$s, while panels (c-f) depict experimental implementations with a fixed time period of 50 $\mu$s. (a) Time evolution of speed with respect to various time periods $\tau$, where dashed and solid lines, respectively, represent $ V_{\mathrm{QSL}}^{s=0}$ and $ V_{\mathrm{QSL}}^{s=-\infty}$. Cases of time periods are colored differently.
		(b) Time evolution of cost rate $\partial_tC$ versus various time periods $\tau$, where cases of time periods are colored differently.
		(c) Applied 20-step sequences (with each of 2.5 $\mu$s) for the laser detuning $\Delta$, the effective Rabi frequency strength $\widetilde{\Omega}$ and the effective phase $\widetilde{\varphi}$. (d) Time evolution of the populations $P_e^x$, $P_e^y$ and $P_e^z$ as defined in \cite{SM}. (e) Time evolution of the actual dynamic speed $|\dot{P}_t|$ in comparison to the two cases of QSL, i.e., $V_{\mathrm{QSL}}^{s=0}$ and $V_{\mathrm{QSL}}^{s=-\infty}$. (f) Time evolution of the actual trade-off ${|\dot{P}_t|}/{\partial_tC}$ in comparison to ${V_{\mathrm{QSL}}^{s=0}}/{\partial_tC}$ and ${V_{\mathrm{QSL}}^{s=-\infty}}/{\partial_tC}$ within the LZ crossing time window. Dots indicate experimental results, while solid lines represent theoretical predictions. Error bars, reflecting the statistical standard deviation, are derived from 50,000 measurements for each data point.}
	\label{Fig2}
\end{figure*}

In the remainder of this Letter, we will verify experimentally the proposed trade-off in the ion-trap platform, using the LZ model under various initial states. Our experiment is carried out with a single ultracold $^{40}$Ca$^+$ ion confined in a linear Paul trap, see Fig. \ref{Fig1}(a), whose axial and radial frequencies are, respectively, $\omega_z/2\pi=1.01$ MHz and $\omega_r/2\pi=1.2$ MHz under the pseudo-potential approximation. The quantization axis is defined by a magnetic field of approximately 6 Gauss at the trap center, which is with respect to the axial direction at an angle $45^\circ$. Prior to the experiment, we have cooled the ion down to near the ground state of the vibrational modes by Doppler and resolved sideband cooling, which is sufficient to avoid detrimental effects of vibrational modes on our operations.

\begin{figure}[t]
	\centering
	\includegraphics[width=8.7cm]{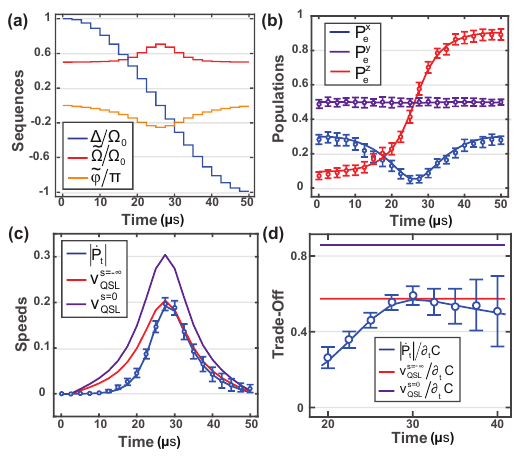}
	\caption{Experimental implementation in the case of initially prepared TES, where the time period is fixed at 50 $\mu$s. The panels (a, b, c, d) here correspond, respectively, to (c, d, e, f) in the previous figure. The characteristic parameters are the same as defined in the previous figure. }
	\label{Fig3}
\end{figure}

As plotted in Fig.~\ref{Fig1}(b), we encode the qubit in electronic states $|4^2S_{1/2},m_{J}=+1/2\rangle$ (labeled as $|g\rangle$) and $|3^2D_{5/2},m_{J}=+5/2 \rangle$ (labeled as $|e\rangle$) with $m_{J}$ the magnetic quantum number, and the qubit is manipulated by an ultra-stable narrow linewidth 729-nm laser with the Lamb-Dicke parameter about $ 0.1$.  As elucidated below, we will consider experimentally both the initially prepared pure state and the initially prepared thermal equilibrium state (TES), i.e., $\rho^{\rm eg}(0) = \operatorname{exp}(-\beta H_{0}(0))/Z_0$ with the partition function $Z_0=\operatorname{tr}[\operatorname{exp}(-\beta H_{0}(0))]$ and $\beta=1/{k_B T}$ (where $T$ denotes the temperature and $k_B$ is Boltzmann constant). For the latter case, we need to introduce an extra energy level $|4^2P_{3/2}, m_J = +3/2 \rangle$ (labeled as $|p\rangle$) for initial state preparation. $|p\rangle$ couples to $|e\rangle$ by a 854-nm laser (with Rabi frequency $\Omega_1$) and dissipates to $|g\rangle$ with the decay rate of $\Gamma_e/2\pi = 23.1$ MHz. Using this dissipative channel, we can rapidly acquire the required TES via reinforcement learning \cite{SM},

Figure~\ref{Fig1}(c) plots the scheme of our experimental steps for implementing a LZ model. Here the LZ model is described by $H_0(t) = \Delta\sigma_x + g(t)\sigma_z$,
where $\Delta$ is the energy splitting, $g(t)$ is the time-dependent field, and $\sigma_{ i\in(x,z)}$ is the Pauli operator.
The instantaneous eigenstates of LZ are $\left|\psi_{-}(t)\right\rangle=\sin (\theta(t))|g\rangle-\cos (\theta(t))|e\rangle$ and  $|\psi_+(t)\rangle = \cos(\theta(t))|g\rangle+\sin(\theta(t))|e\rangle$ with $\cos(2\theta(t))=g(t)/\sqrt{g^2(t)+\Delta}$ and $\sin(2\theta(t))=\Delta/\sqrt{g^2(t)+\Delta}$.
According to Ref. \cite{JPA2009}, the corresponding ancillary counterdiabatic term is given by
 \begin{equation}
  	H_{1}(t) = -\frac{\hbar g^{\prime}(t)\Delta}{2(\Delta^2+g^2(t))}\sigma_y,
  	\label{Eq9}
 \end{equation}
which allows us to investigate Eqs. (\ref{Eq4}) and (\ref{Eq7}).
Thus, the combined Hamiltonian is $H(t)=H_0(t) + H_{1}(t)$ and can be rewritten as
\begin{equation}
	H(t) = \widetilde{\Omega}(t)(e^{\rm i\tilde {\varphi}(t)}\sigma_+ + e^{\rm -i\tilde{\varphi}(t)}\sigma_-) + g(t)\sigma_z.
	\label{Eq10}
\end{equation}
Equation (\ref{Eq10}) is a modified form of the LZ model with the effective Rabi frequency $\widetilde{\Omega}(t) = \sqrt{4\Delta^2 +(\Xi)^2 }$ and the phase $\widetilde{\varphi}(t)=\arctan(\Xi/(2\Delta))$, where $\Xi=-\frac{\hbar g^{\prime}(t)\Delta}{2(\Delta^2+g^2(t))}$.
So our LZ model can be carried out by exactly controlling the irradiation power and the phase of 729-nm laser, under the government of Eq. (\ref{Eq10}).

Here we first verify the proposed trade-off from the initially prepared pure state, i.e., an eigenstate $|\psi_+(0)\rangle$.
In this case, with the counterdiabatic field, i.e., Eq.~(\ref{Eq9}) applied, the system evolves around the instantaneous eigenstate $|\psi_+(t)\rangle$. For the condition far from the adiabatic limit, we choose $\Delta=\Omega_{0}/4$ and $g(t)=(\Omega_0 /2)\cos(\pi t/\tau)$ with $\Omega_0/2\pi=40$ kHz, and consider various time periods $\tau$.
For an intuitive comparison of QSL and cost rate, we set $\hbar=1$. Figures~\ref{Fig2}(a,b) plot the theoretical simulation of the QSL based on Eqs.~(\ref{Eq2}) and (\ref{Eq4}), including ${V_{\mathrm{QSL}}^{s}}$ with Wigner phase space $s=0$ or $s=-\infty$ and the cost rate ${\partial_tC}$.
It is evident that $V_{\mathrm{QSL}}^{s=0}$ is less accurate for shorter time period, indicating that $V_{\mathrm{QSL}}^{s=-\infty}$ is better for describing the QSL in the case of short time periods.
Comparing Fig.~\ref{Fig2}(a) and Fig.~\ref{Fig2}(b), we see that the cost rate varies  consistently with the speed, showing that faster manipulation requires higher energy consumption.

Experimentally, we focus on the case of $\tau=50~\mu$s. The initial state $|\psi_+(0)\rangle$ is prepared by applying a carrier-transition pulse with $\theta=0.8524\pi$ and $\tilde \varphi=\pi/2$. Variation of the characteristic parameters of the LZ model is made by operational sequences with 20 steps, as sketched in Fig. \ref{Fig2}(c). The powers, frequencies and phases of the lasers are controlled via acousto-optic modulators and an arbitrary waveform generator. After applying the laser pulses, we measure the population in the excited state $|e\rangle$ in three directions, labeled as $P_e^{i\in(x,y,z)}$, which constructs the quantum tomographic results of $\rho_t$ \cite{SM}, see Fig.~\ref{Fig2}(d).

To examine the optimal QSL, we compare the results from Eq. (\ref{Eq4}) to the actual dynamical speed of the quantum state, i.e., $|\dot{P}_t|= |\operatorname{tr}(\rho_0\dot{\rho}_t)| = \frac{1}{\hbar }|\operatorname{tr}(\rho_0H(t)\rho_t)-\operatorname{tr}(\rho_0\rho_tH(t))|$ with $\rho_0 = |\psi_+(0)\rangle\langle\psi_+(0)|$. Figure \ref{Fig2}(e) presents the experimentally measured dynamical speed $|\dot{P}_t|$, facilitating a comparison with the QSL employing various phase space parameters, such as $s=-\infty$ and $s=0$ (Wigner phase space). We find that, although both $V_{\rm QSL}^{s=-\infty}$ and $V_{\rm QSL}^{s=0}$ have maximal values at  $t=27.5$ $\mu$s, varying consistently with $|\dot{P}_t|$, the former is tight at $t=29$ $\mu$s (see proof in Ref. \cite{SM}), indicating that $V_{\rm QSL}^{s=-\infty}$ represents the optimal QSL we acquired. This optimality hinges on two factors: one being the tightness of the bound, and the other its ability to accurately reflect the actual dynamical speed.

Based on the optimal QSL, we are ready to examine the optimal trade-off, i.e., Eq. (\ref{Eq7}), as illustrated in Fig. \ref{Fig2}(f). Note that,
although the upper bound of $V_{\rm QSL}^{s=-\infty}$ matches very well the experimentally actual dynamical speed, the deviation between the two curves is evident in the regime of $t< 29~\mu$s, see Fig. \ref{Fig2}(e). We emphasize that the difference originates from the fact that the QSL  curves are nearly symmetric whereas the curve of the experimental actual dynamical speed is not \cite{explain-1}.
Besides, for the experimental data, the error bars of the data points after t$=30~\mu$s become larger and larger with time, which reflect more imprecision and uncertainty involved in the measurement. We conjecture that the enlarged uncertainty  comes from the emergence of the complex dynamics when the state evolves through the avoided crossing area (around 29 $\mu$s) of the LZ model. This uncertainty is amplified  in Fig.~\ref{Fig2}(f) by the cost rate ${\partial_tC}$ whose values are rapidly shrinking with $t > 29~\mu$s. Consequently, to verify the optimal trade-off more accurately, it is highly expected to detect the spin states with ultimate precision.

Next, we consider a more complex situation, verifying the trade-off in the case of an initially prepared mixed state, i.e., a TES. For simplicity of our experimental implementation, we fix the inverse temperature $\beta \Omega_0= 5\times 10^{28}$.
Similar to the above case for initially prepared pure state, we design the operational sequences with 20 steps to vary the characteristic parameters of the LZ model, as sketched in Fig.~\ref{Fig3}(a), and with the laser pulses we measure the quantum tomographic results of $\rho_t$, see Fig.~\ref{Fig3}(b). Differently, the initial state prepared in this case is more complicated. We switch on the dissipative channel (i.e., $|e\rangle\rightarrow|p\rangle\rightarrow|g\rangle$) and employ reinforcement learning to prepared the required TES \cite{SM}.
The test of our proposed trade-off is executed similarly to the above case of initially pure state. From the observations in Figs.~\ref{Fig3}(c,d), we may draw the same conclusions as in the case of the initially pure state that $V_{\rm QSL}^{s=-\infty}$ represents the optimal upper bound of the QSL and thus reaches the optimal trade-off.

In summary, we have established an optimal trade-off between the dynamical speed and the cost rate of the transitionless driving field in the context of shortcuts to adiabaticity, utilizing the $s$-parametrized phase space technique. We have justified both theoretically and experimentally that this novel trade-off stands out from previous relations by being tight. These findings substantially enhance our comprehension of the fundamental constraints inherent in quantum operations, underscoring the impossibility of arbitrarily fast processes. Moreover, the method and results of our work promote the understanding that less commonly utilized phase spaces, which have been largely overlooked, may offer superior advantages in certain quantum tasks.

We appreciate the discussion with Qingshou Tan for preparing initial mixed state. This work was supported by National Natural Science Foundation of China under Grant Nos. 12304315, U21A20434, 12074346, 12074390, 12074280, 92265107, by China Postdoctoral Science Foundation under Grant Nos. 2022M710881, 2023T160144, by Key Lab of Guangzhou for Quantum Precision Measurement under Grant No. 202201000010, by Science and Technology Projects in Guangzhou under Grant Nos. 202201011727 and 2023A04J0050, and by Nansha Senior Leading Talent Team Technology Project under Grant No. 2021CXTD02.

\bibliography{references_QSL}

\begin{thebibliography}{99}
\bibitem{Note} While a large number of metrics exist, the Bures fidelity often emerges as one of the preferred choices. Nevertheless, the QSL bound, defined in terms of the Bures angle $\mathcal{L}_t=\arccos \left|\left\langle\psi_0 \mid \psi_t\right\rangle\right|$ as outlined in Ref. \cite{PRLSS}, may not only exhibit a lack of tightness but also potentially contradict the true dynamics, as detailed in \cite{SM}. Furthermore, the computation of Bures angle for mixed states is usually very challenging. In this Letter, we will employ the relative purity \cite{QICAM,PRLAI,Lidar98,PRLBA} instead.
\bibitem{explain} For the case of initially mixed states, it is theoretically cumbersome to establish a concise functional relationship between the speed limit and the cost rate, i.e., $V_{\mathrm{QSL}}^{-\infty}[\partial_t C]$. Consequently, the ratio $V_{\mathrm{QSL}}^{-\infty}(t)/\partial_t C$ continues to be utilized, as it can be directly assessed through numerical simulations based on Eqs. (\ref{Eq2a}) and (\ref{Eq4}).
\bibitem{explain-1} For the deviation of the actual dynamic speed from the curve of $V_{\rm QSL}^{s=-\infty}$, we emphasize that the tight QSL bound and the actual dynamical speed might not coincide at the midpoint of the driving interval. The coincidence time point between the two curves completely relies on the driving Hamiltonian (e.g., the $g(t)$ function in the LZ model) and the actual manipulation (e.g., the operational sequences) involved in our experiment. In our case here, the QSL bounds for the LZ model, i.e., the purple and red curves, look nearly symmetrical, wherea the actual dynamical speed exhibits evident asymmetry before and after $t= 29~\mu$s. This asymmetry primarily accounts for the noticeable discrepancy between the QSL bound and the actual dynamical speed observed over the short time period. This clarification applies to Fig. 3(c).
\end{thebibliography}

\pagebreak
\widetext
\begin{center}
\vspace{1cm}
\textbf{{\large Supplemental Material}}
\end{center}

\setcounter{equation}{0} \setcounter{figure}{0} \setcounter{table}{0}
\makeatletter
\renewcommand{\theequation}{S\arabic{equation}} \renewcommand{\thefigure}{SM%
\arabic{figure}} \renewcommand{\bibnumfmt}[1]{[#1]} \renewcommand{%
\citenumfont}[1]{#1}

\tableofcontents

\vspace{1cm}
We first discuss analytically some theoretical details about the metric and quantum speed limit (QSL) associated with our experiment. Then we present some details for the state tomography and initial state preparation in our experiment.

\section{I. Advantages of the quantum speed limit utilized in this study} \label{Sec-Metric}
A seminal trade-off is established between the upper bound of dynamical speed, known as the quantum speed limit (QSL), and the cost rate associated with the auxiliary driving Hamiltonian \cite{PRLSS}. Notably, while much attention has been directed towards the QSL, the actual dynamical speed often remains overlooked. We argue that two factors are crucial in evaluating the efficacy of this trade-off: firstly, the tightness of the bound, and secondly, the extent to which the bound accurately reflects the true dynamics.

\begin{figure}[tb]
	\centering
	\includegraphics[width=18.0cm]{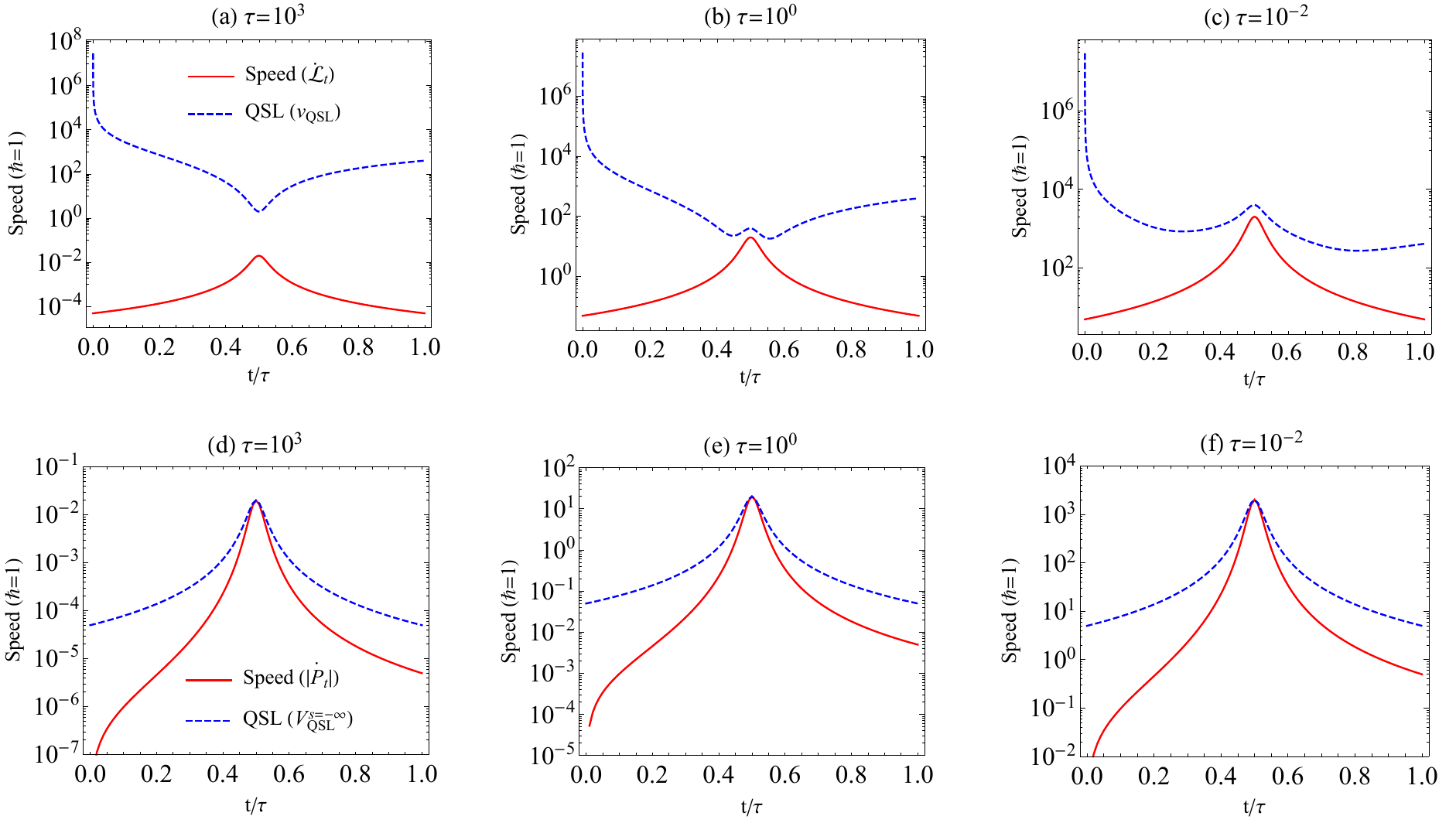}
	\caption{Dynamical speed of a qubit and its associated quantum speed limit (QSL) in a Landau-Zener model. The model is governed by \( H_0(t) = \Delta \sigma_x + g(t) \sigma_z \), with the evolution influenced by an ancillary counterdiabatic field \( H_1(t) = -\Delta g^{\prime}(t) \sigma_y / [2(\Delta^2 + g^2(t))] \) within $t \in [0, \tau]$. A linear ramp is defined as \( g(t) = 0.2 - 0.4(t / \tau) \). Panels (a)-(c) utilize the metric as described in Ref. \cite{PRLSS}, whereas panels (d)-(f) implement the metric enhanced by the $s$-parameterized phase space technique from Ref. \cite{PRAWZ}. For these simulations, \( \Delta = 0.01 \).}
	\label{figmetrics}
\end{figure}

In Ref. \cite{PRLSS}, the measure of dynamical speed is characterized by the change rate of the angle $\mathcal L_t=\arccos \left|\left\langle\psi_0 \mid\psi_t\right\rangle\right|$ between the initial state $\left|\psi_0\right\rangle$ and the evolved state $\left|\psi_t\right\rangle$. This speed is expressed as
\begin{equation}
	\text{Speed}\to\dot{\mathcal{L}}_t,
\end{equation}
which is constrained by the QSL
\begin{equation}
	\mathrm{QSL}\to v_{\mathrm{QSL}}=\frac{\sqrt{\epsilon_n^2(t)+\left\langle\partial_t n_t \mid \partial_t n_t\right\rangle}}{\hbar \cos \left(\mathcal{L}_t\right) \sin \left(\mathcal{L}_t\right)}, \label{QSL}
\end{equation}
namely, $\mathcal L_t \leq v_{\mathrm{QSL}}$, where $\epsilon_n(t)$ and $\left| n_t \right\rangle$ are the instantaneous eigenvalues and eigenstates of a time-dependent Hamiltonian $H_0(t)$. However, there exists a significant discrepancy between the dynamical speed and its associated QSL. In certain instances, the QSL bound may not only lack tightness but can also contradict the true dynamics, as exemplified in Fig. \ref{figmetrics}(a)-(c).

To address this issue, we utilize the change rate of the relative purity $P_t\left(\rho_0, \rho_t\right)=\operatorname{tr}\left(\rho_0 \rho_t\right)$
\begin{equation}
	\text{Speed}\to \left|\dot{P}_t\left(\rho_0, \rho_t\right)\right|
\end{equation}
as a measure for dynamical speed. For pure states, $\mathcal L_t=\arccos\sqrt{P_t}$. Obviously, such relative purity, in isolation, is insufficient to resolve the tightness issue associated with the QSL. However, this limitation can be effectively circumvented through the application of the recently developed $s$-parameterized phase space technique in the context of QSL \cite{PRAWZ}.

The relative purity in $s$-parameterized phase space is expressed as (see details in Ref. \cite{PRAWZ})
\begin{equation}
    P_t\left(\rho_0, \rho_t\right)=\int d \mu(\eta) F_{\rho_0}^{-s}(\eta) F_{\rho_t}^s(\eta), \label{purityP}
\end{equation}
where $F_A^s(\eta)$ is the phase space function of the operator $A$, $\eta$ is a point in a phase space that determines a state $|\eta\rangle~(\eta \rightarrow|\eta\rangle)$ in Hilbert space, and the index $s$ labels a family of phase spaces. The time derivative of Eq. (\ref{purityP}) yields
\begin{equation}
    \dot{P}_t\left(\rho_0, \rho_t\right)=\int d \mu(\eta) F_{\rho_0}^{-s}(\eta)\left\{\left\{F_H^s, F_{\rho_t}^s\right\}\right\}(\eta), \label{Ptt}
\end{equation}
where we have used the von Neumann equation in phase space, i.e., $\frac{\partial F_{\rho_t}^s(\eta)}{\partial t}=\left\{\left\{F_H^s, F_{\rho_t}^s\right\}\right\}(\eta)$. Here $\{\{\cdot,\cdot\}\}$ denotes as the generalized Moyal bracket. Using the Cauchy-Schwarz inequality, the corresponding QSL is defined as
\begin{equation}
	\mathrm{QSL}\to \sup _{s \in S} V_{\mathrm{QSL}}^s(t), \label{VQSLs}
\end{equation}
namely, $\left|\dot{P}_t\left(\rho_0, \rho_t\right)\right| \leq \sup _{s \in S} V_{\mathrm{QSL}}^s(t)$, with
\begin{equation}
    V_{\mathrm{QSL}}^s(t)=\left[\int d \mu(\eta) F_{\rho_0}^{-s}(\eta)^2\right]^{\frac{1}{2}}\left[\int d \mu(\eta)\left|\left\{\left\{F_{\rho_t}^s, F_H^s\right\}\right\}(\eta)\right|^2\right]^{\frac{1}{2}}.
\end{equation}
The selection of the parameter $s$ in a set $S$ hinges on the specific model in question. For instance, in a qubit system, $S=\mathbbm{R}$ and the least upper bound is attained at $s=-\infty$ (see proof in Sec. II).

We now proceed to demonstrate the superiority of Eq. (\ref{VQSLs}) over Eq. (\ref{QSL}) using the Landau-Zener model as an example. As illustrated in Fig. \ref{figmetrics}(d)-(f), the QSL bound is tight (see proof in Sec. II) and closely aligns with the actual dynamical speed, providing a more accurate representation than what is observed in Fig. \ref{figmetrics}(a)-(c). Moreover, the metric used in this Letter is applicable to any initial state, pure or mixed.

\section{II. The tightness of the quantum speed limit} \label{Sec-Tight}
In the main text, a concise and rigorous proof of the QSL bound's tightness for Eq. (\ref{VQSLs}) is presented. This section outlines the proof and illustrates, through two examples, the application of the tightness condition specified in Eq. (3) of the Letter.

According to Eq. (\ref{Ptt}), the condition for equality in $\left|\dot{P}_t\left(\rho_0, \rho_t\right)\right| \leq \sup _{s \in S} V_{\mathrm{QSL}}^s(t)$ is met when $F_{\rho_0}^{-s}(\eta)$ and $\dot F_{\rho_t}^{s}(\eta)=\left\{\left\{F_H^s, F_{\rho_t}^s\right\}\right\}(\eta)$ exhibit linear relevance. Without loss of generality, let us consider $N$-level quantum systems. After some algebra \cite{PRAWZ}, we have
\begin{equation}
	F_{\rho_0}^{-s}(\eta)=\frac{1}{N}+2 b_{\mu}(0) r_{-s} R_{\mu},\label{lhsF}
\end{equation}
and
\begin{equation}
	\left\{\left\{F_H^s, F_{\rho_t}^s\right\}\right\}(\eta)=\frac{2}{\hbar}r_s b_{\mu}(t)h_{\nu}(t)R_{\lambda}f_{\nu \mu \lambda},\label{rhsF}
\end{equation}
where $r_s=\frac{1}{2} \sqrt{(N+1)^{1+s}}$ ($s\in\mathbbm{R}$), $b_\mu(t)=2 \operatorname{tr}\left(\rho_t T_\mu\right)$, $h_\mu(t)=2 \operatorname{tr}\left(H(t) T_\mu\right)$, and $R_\mu=\left\langle\eta\left|T_\mu\right| \eta\right\rangle$. $T_\nu$ are the generators of SU(N) Lie algebra and $f_{\beta \gamma \xi}$ are totally antisymmetric regarding the interchange of any pair of its indices (see details in Ref. \cite{PRAWZ}). As $s$ approaches $-\infty$, the value of $r_{-s}$ tends towards infinity. Consequently, the first term, $1/N$, in Eq. (\ref{lhsF}) becomes negligible and can be omitted. We can solve the following set of simultaneous equations to ensure linear relevance between Eqs. (\ref{lhsF}) and (\ref{rhsF})

\begin{equation}
	b_{\mu}(0) \propto h_{\nu}(t) b_{\lambda}(t)f_{\nu \lambda \mu}.\label{TightConditions}
\end{equation}
Generally, Eqs. (\ref{TightConditions}) lend themselves to numerical solutions, except in certain special cases where an analytical result can be obtained. In the following, we consider two examples.

\begin{figure}[t]
	\centering
	\includegraphics[width=7.0cm]{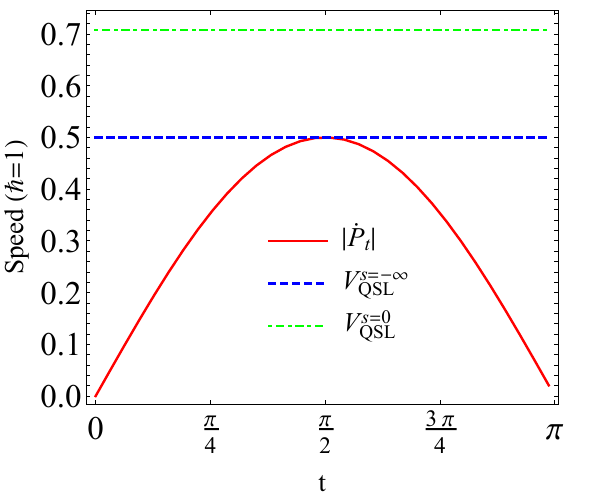}
	\caption{Dynamical speed of a qubit and its associated quantum speed limit bounds. The model is governed by $H=\frac{\hbar}{2} \sigma_z$ with an initial state prepared in $(|0\rangle+|1\rangle)/\sqrt{2}$. The bound is achieved when $s\to \infty$ and $t=\pi/2$, which implies that $V_{\mathrm{QSL}}^{-\infty}$ is tight.}
	\label{figExample1}
\end{figure}

\textit{\textbf{Example 1}}-Spin-$\frac{1}{2}$ rotations. Consider a qubit initially prepared in state $\rho_0=|\phi\rangle \langle\phi|$ with $|\phi\rangle=(|0\rangle+|1\rangle)/\sqrt{2}$, and driven by the Hamiltonian $H=\frac{\hbar}{2} \sigma_z$. Thus, $h_1 =h_2 =0$, and $h_3=\hbar$. The evolved state is given by $\rho_t=\frac{\mathbbm{1}}{2}+\cos{t}\sigma_x+\sin{t}\sigma_y$, which yields $b_1=2 \cos{t}$, $b_2=2\sin{t}$, and $b_3=0$. According to Eqs. (\ref{TightConditions}), we have
\begin{equation}
	\sin{t}=\pm 1 \: (\text{or} \; \cos{t}=0),
\end{equation}
which implies that when $s\to -\infty$ and $t=\pi/2$ (or strictly $\pi(l-1/2)$, $l \in \mathbbm{Z}^+$), the upper bound in Eq. (\ref{VQSLs}) is tight (see Fig. \ref{figExample1}).

\textit{\textbf{Example 2}}-Landau-Zener model. Consider a qubit initially prepared in one of the eigenstate of $H_0(0)=\Delta \sigma_x+g(0)\sigma_z$. The system is then driven by the total Hamiltonian $H(t)=H_0(t)+H_1(t)$, with the ancillary counterdiabatic field $-\Delta g^{\prime}(t) \sigma_y /\left[2\left(\Delta^2+g^2(t)\right)\right]$. After some algebra, Eqs. (\ref{TightConditions}) yield
\begin{equation}
	b_3(t) h_2(t)-b_2(t) h_3(t) = \frac{b_1(0)}{b_3(0)}[b_2(t) h_1(t)-b_1(t) h_2(t)], \label{Exemple2-a}
\end{equation}
or
\begin{equation}
	b_1(t) h_3(t)-b_3(t) h_1(t) =0.\label{Exemple2-b}
\end{equation}

For example, let us consider the parameters utilized in Sec. \ref{Sec-Metric}, i.e., $\Delta=0.01$ and $g(t)=0.2-0.4 t/\tau$. Numerical results for Eq. (\ref{Exemple2-a}) or Eq. (\ref{Exemple2-b}) were obtained at $t/\tau=0.501$ ($\tau=10^3, 10^0$, and $10^{-2}$), as illustrated in Fig. \ref{figmetrics}(d)-(f). For experimental data in the main text, i.e., $g(t)=\frac{1}{2}\hbar \Omega_0 \cos(\pi t/\tau)$, $\tau=50$ $\mu s$, $\Delta=\frac{1}{4}\hbar \Omega_0$, and $\Omega_0/{2\pi}=40$ kHz, Eq. (\ref{Exemple2-a}) or Eq. (\ref{Exemple2-b}) were solved numerically at $t=29$ $ \mu s$.

\section{III. Tomography of a single qubit state }
An arbitrary state of a single qubit can be uniquely represented as
\begin{equation}
	\rho=\frac{1}{2} \left( I+ \sum_{i=x,y,z }S_i \sigma_i \right),
\end{equation}
which is formed by Pauli operators $\sigma_i$ and Stokes parameters $S_i$.
We have $S_i= \rm Tr[\sigma_i\rho]$, and these parameters as below need to be measured.
\begin{equation}
	S_x=P_{e}^x-P_{g}^x, \quad S_y=P_{e}^y-P_{g}^y, \quad S_z=P_{e}^z-P_{g}^z,
\end{equation}
where P$^x$ is a projection measurement on $|g\rangle^x=\frac{1}{\sqrt{2}}(|e\rangle-|g\rangle)$ and $|e\rangle^x=\frac{1}{\sqrt{2}}(|e\rangle+|g\rangle)$. P$^y$ is a projection measurement on $|g\rangle^y=\frac{1}{\sqrt{2}}(|e\rangle-i|g\rangle)$ and $|e\rangle^y=\frac{1}{\sqrt{2}}(|e\rangle+i|g\rangle)$.
P$^z$ is a projection measurement on $|g\rangle$ and $|e\rangle$.
Due to $P_e^i+P_g^i=1~({i\in(x,y,z)})$, $S_i$ can be rewritten as $S_x = 2P_{e}^x-1$, $\quad S_y = 2P_{e}^y-1$ and $\quad S_z = 2P_{e}^z-1$.
Then the tomography of the single qubit state is executed via the steps described below.

\noindent\textbf{\uppercase\expandafter{(\romannumeral1)} Measurement of $S_z$}: We measure $\rho$ in $|e\rangle$ to obtain $P_e$. $S_z$ is calculated by $S_z=2P_e^z-1$.

\noindent\textbf{\uppercase\expandafter{(\romannumeral2)}  Measurement of $S_x$}: We rotate $\rho$ by applying a carrier-transition pulse with $\theta=\phi=\pi/2$, following a measurement of $\rho$ in $|e\rangle$ to obtain $P^x_e$. $S_x$ is thereby calculated by $S_x=2P^x_e-1$.

\noindent\textbf{\uppercase\expandafter{(\romannumeral3)}  Measurement of $S_y$}: We rotate $\rho$ by applying a carrier-transition pulse with $\theta=\pi/2, \phi=0$, following a measurement of $\rho$ in $|e\rangle$ to obtain $P^y_e$. $S_y$ is then calculated by $S_y=2P^y_e-1$.
\\

\section{IV. Preparing the thermal equilibrium state via reinforcement learning }
\begin{figure}[htbp]
	\centering
	\includegraphics[width=18 cm]{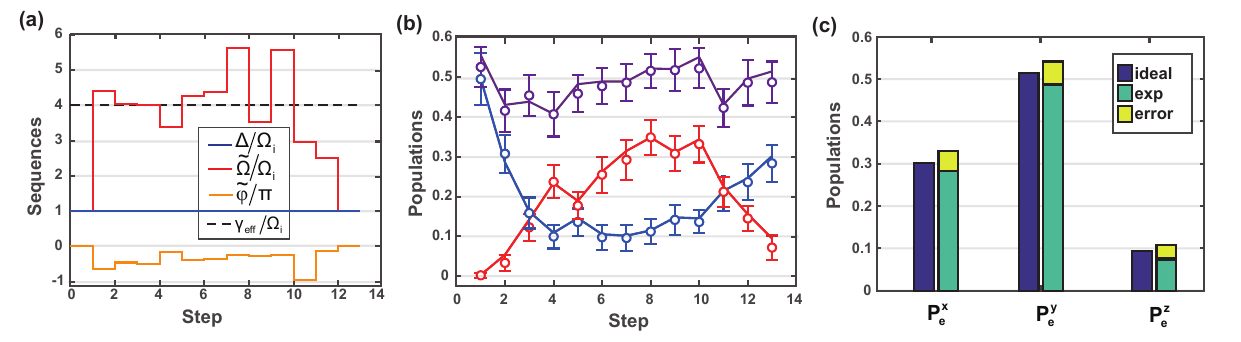}
	\caption{ (a) Designed pulses of the detuning, the Rabi frequency strength, the phase and the effective decay for initial state preparation via RL method. Here, we choose $\Omega_i/2\pi = 20$ kHz. (b) Time evolution of the populations $P^x_e$, $P^y_e$ and $P^z_e$, corresponding, respectively, to the blue, purple, and red lines. Here, the dots are	experimental results and lines represent theoretical result by simulating the master equation. The error bars indicating the statistical standard deviation of the experimental data are obtained by 10,000 measurements for each data point. (c) The populations $P^x_e$, $P^y_e$ and $P^z_e$  of the last step (the green bars, exp), and compare them with the theoretical simulation (the blue bars, ideal). The errors (yellow bars) are from the experimental statistical errors. }
	\label{Fig.S1}
\end{figure}

In our experiment, the initial thermal equilibrium state $\rho^{\rm eq}(0)$, mentioned in main text, is prepared by reinforcement learning (RL).
Concretely, we first cool the trapped ion down to near the ground vibrational state and then prepare the state $|g\rangle$ by optical pumping.
Then, the initial state is prepared by the following Hamiltonian,
\begin{equation}
	H(t) = \frac{\delta}{2}\sigma_z + \frac{\Omega_i}{2}\sigma_x + 2f_{\rm opt1}(t)\sigma_y + 2f_{\rm opt1}(t)\sigma_x,
\end{equation}
where $\Omega_i$ represents the Rabi frequence in the case of state preparation, $f_{\rm opt1}(t)\sigma_y$ and $f_{\rm opt2}(t)\sigma_x$ are the control terms imposed by the RL approach.
Experimentally, this Hamiltonian can be carried  out by using a single beam of 729-nm laser  with
time-dependent Rabi frequency  $\tilde{\Omega}(t)=\sqrt{16f_{\rm opt2}^2+(\Omega_i+4f_{\rm opt2})^2}$, phase $\phi(t)=\arctan(-4f_{\rm opt1}/(\Omega_i+4f_{\rm opt2}))$, and laser detuning $\delta$.
Meanwhile,   the 854-nm laser is switched on  to construct  the dissipative channel.
Therefore, the whole system is governed by the Lindblad master equation,
\begin{equation}
	\dot{\rho} =-i[H, \rho] + \frac{\gamma_{\rm eff}}{2}(2 \sigma_{-}\rho\sigma_{+} -\sigma_{+}\sigma_{-}\rho - \rho\sigma_{+}\sigma_{-} ),
	\label{Eq2}
\end{equation}
where $\rho$ denotes the density operator, $\gamma_{\rm eff} = \Omega_1^2/\Gamma$ is the effective decay, $\sigma_{-}=(\sigma_{+})^{\dagger} \equiv|g\rangle \langle e|$ is the usual Pauli operator, and the level scheme is shown in Fig. 1(a) in the main text.

With the RL-designed optimal pulses (see Fig. \ref{Fig.S1}(a)), after thirteen steps (taking 13 $\mu$s) with each step of the length $0.125/\Omega_i$, the system  is prepared to $\rho^{\rm eq}(0)$.
Experimentally, we measure the time evolution of the populations in comparison with the theoretical simulation, as plotted in Fig. \ref{Fig.S1}(b). At the final step, we find the fidelity $F = 99.31\%$, as shown in Fig. \ref{Fig.S1}(c).

\end{document}